\documentclass[prb,notitlepage]{revtex4-1} 
\usepackage{graphicx} 
\usepackage[margin=0.7in]{geometry}
\usepackage{amsmath}
\usepackage{multirow}
\usepackage{newtxtext}                                        
\usepackage{booktabs}
\usepackage[dvipsnames]{xcolor}
\usepackage[normalem]{ulem}
\usepackage[textwidth=1.5cm,textsize=footnotesize]{todonotes}
\usepackage[cmintegrals]{newtxmath}
\newcommand{\bvec}[1]{\boldsymbol{\mathrm{#1}}}
\usepackage{lipsum}
\usepackage{float}

\newcommand{\BostonCollege}{Department of Physics, Boston College, Chestnut Hill, MA, USA}
\newcommand{\HarvardSEAS}{John A. Paulson School of Engineering and Applied Sciences, Harvard University, Cambridge, MA, USA}
\newcommand{\MaxPlanck}{Max Planck Institute for Chemical Physics of Solids, Dresden, Germany}

\begin{document}
\author{Gavin B. Osterhoudt}\affiliation{\BostonCollege}
\author{Yaxian Wang}\affiliation{\HarvardSEAS}
\author{Christina A.C. Garcia}\affiliation{\HarvardSEAS}
\author{Vincent M. Plisson}\affiliation{\BostonCollege}
\author{Johannes Gooth}\affiliation{\MaxPlanck}
\author{Claudia Felser}\affiliation{\MaxPlanck}
\author{Prineha Narang}\affiliation{\HarvardSEAS}
\author{Kenneth S. Burch}\email{ks.burch@bc.edu}\affiliation{\BostonCollege}

\title{Evidence for dominant phonon-electron scattering in Weyl semimetal WP$_{2}$}

\begin{abstract}
Topological semimetals have revealed a wide array of novel transport phenomena, including electron hydrodynamics, quantum field theoretic anomalies, and extreme magnetoresistances and mobilities. However, the scattering mechanisms central to these behaviors remain largely unexplored. Here we reveal signatures of significant phonon-electron scattering in the type-II Weyl semimetal WP$_{2}$ via temperature dependent Raman spectroscopy. Over a large temperature range, we find that the decay rates of the lowest energy $A_{1}$ modes are dominated by phonon-electron rather than phonon-phonon scattering. In conjunction with first-principles calculations, a combined analysis of the momentum, energy, and symmetry-allowed decay paths indicates this results from intraband scattering of the electrons. The excellent agreement with theory further suggests that such results could be true for the acoustic modes. We thus provide evidence for the importance of phonons in the transport properties of topological semimetals and identify specific properties that may contribute to such behavior in other materials.
\end{abstract}

\maketitle

Topological semimetals display a range of novel transport phenomena, including enormous magnetoresistance and mobilities\cite{shekharNPhys2015,wangPRB2016,kumarNComm2017,aliNat2014}. Observations of such behavior in topological systems suggested the quantum geometry of the electronic bands was a crucial ingredient. However, non-topological semimetals such as LaAs or PtSn$_{4}$ have revealed similarly remarkable magnetoresistances and mobilities\cite{yangPRB2017,munPRB2012,duPRB2018}, suggesting that these properties are more endemic to semimetals in general, with some proposing the primary factor is the near-perfect electron-hole compensation\cite{yangPRB2017}. In addition, there is often enormous variation in the temperature dependence of these properties, indicating that temperature dependent scattering processes, such as those provided by electron-phonon coupling, are important.

The transition metal dipnictide WP$_{2}$ is an ideal semimetal in which to study the role of coupling between the electron and phonon systems. The topological phase, $\beta-$WP$_{2}$, displays the largest magnetoresistance of any topological semimetal\cite{kumarNComm2017}, with hints of hydrodynamic behavior at low temperatures\cite{goothNComm2018}. In WP$_{2}$ the resistivity decreases, while the mobility increases, by four and five orders of magnitude respectively between room temperature and 2~K\cite{kumarNComm2017}. Over the temperature range where the majority of this change occurs the resistivity appears to be dominated by electron-phonon scattering\cite{kumarNComm2017,jaouinpjQ2018}. Previous computational works also suggested that electron-phonon coupling plays a large role in determining the macroscopic transport properties of WP$_{2}$\cite{coulterPRB2018}.

An important and related consideration is the phonon-electron scattering, i.e. the scattering of phonons by electrons. Phonon-electron scattering rates are a critical ingredient in the formation of a combined electron and phonon ``fluid'' which precipitates hydrodynamic behavior\cite{levchenko2020,varnavides2020generalized}. Such scattering is of considerable import when evaluating the lattice's effect on transport properties. For example, if phonon-electron scattering rates exceed phonon-phonon rates, momentum lost to phonons may be returned to the charge carriers, leading to an enhanced conductance\cite{peierls1932,Wiser1984,varnavides2020generalized}. Evidence for strong phonon-electron coupling in topological semimetals (TSM) has been primarily reported via optical spectroscopies that directly probe the system's phonons\cite{osterhoudtPRB2019,xuNComm17,zhangPRB2019,sharafeevPRB2017}. However it is unclear what combination of factors contribute to the phonon-electron scattering and thus its role in the transport behavior of TSMs. To this end we performed a combined experimental, computational, and theoretical investigation of the phonon-electron coupling in WP$_{2}$. Using Raman spectroscopy, first principles calculations, and symmetry analysis, we provide evidence that intraband phonon-electron scattering dominates over phonon-phonon scattering over a wide temperature range. The ab-initio calculations further elucidate the roles played by the relative phase space and phonon-electron coupling strength in the dominance of phonon-electron scattering. As such our work guides future efforts to understand and optimize novel transport phenomena in semimetals.

\begin{figure}
\includegraphics{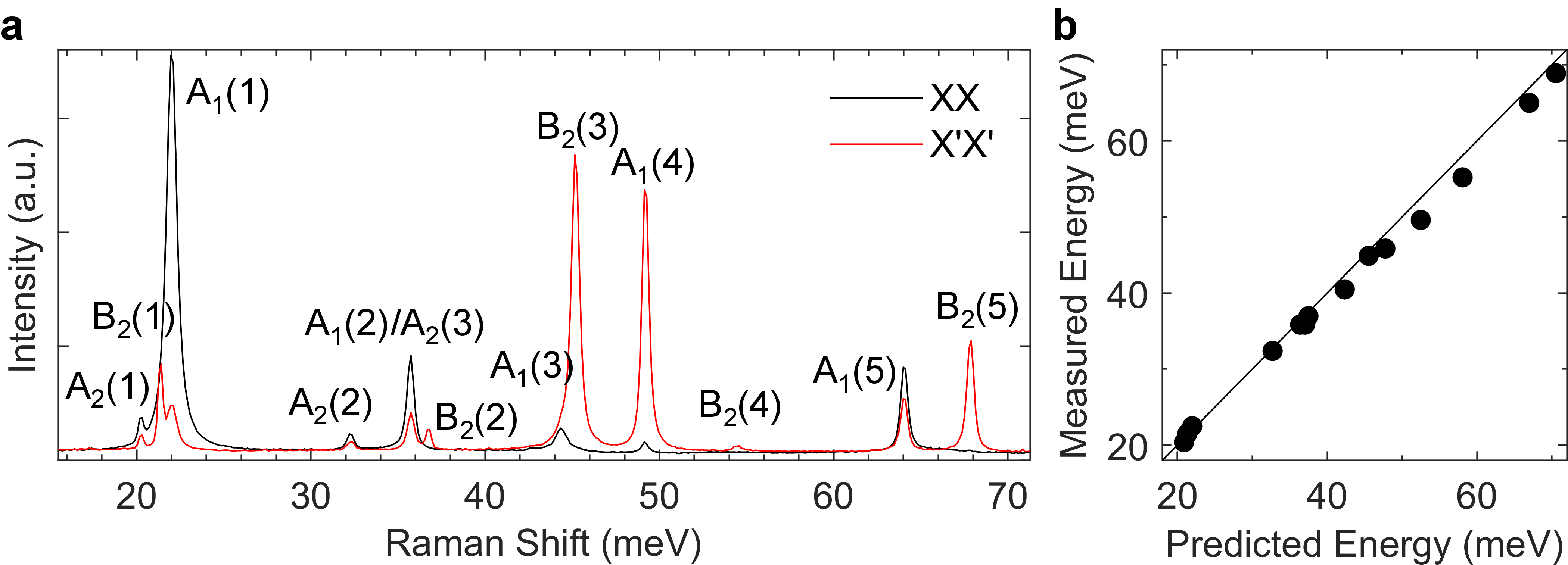}
\caption{(a) Room temperature Raman spectra of an as-grown WP$_{2}$ crystal surface in collinear XX and X'X' polarization configurations. (b) Predicted phonon energies vs measured phonon energies. The solid line indicates ideal agreement.}
\label{Fig1}
\end{figure}

We utilized Raman spectroscopy due to its sensitivity to symmetry, high energy resolution, and the prevalence of phonon-electron coupling signatures in Raman spectra of other TSMs\cite{osterhoudtPRB2019,xuNComm17,zhangPRB2019,sharafeevPRB2017}. As the role of symmetry has so far been largely overlooked, we begin by considering in detail the symmetry properties of WP$_{2}$. The orthorhombic, nonsymmorphic space group $Cmc2_{1}$, leads to a total of 18 phonon modes. The 15 optical modes are all Raman active and a group theoretical analysis\cite{bilbao} gives the irreducible representations 5$A_{1}$, 3$A_{2}$, 2$B_{1}$, and 5$B_{2}$. In Fig.~1a we show the room temperature Raman spectra of an as-grown WP$_{2}$ crystal surface measured in XX and X$'$X$'$ polarization configurations, revealing 14 of the 15 optical modes. The full polarization dependence yields the mode assignments shown in Fig.~1a, which are consistent with previous studies\cite{suAM2019,wulferding2020}(Supplemental Table~1). In Fig.~1b we show our measured versus computationally predicted mode energies. The solid diagonal line indicates ideal agreement, and the reported values all fall within 3\% of this line.

We begin our search for signatures of phonon-electron coupling in the temperature dependence of the phonon energies. Typically the phonon energy decreases as the temperature is raised due to a combination of lattice expansion and anharmonic renormalization\cite{balkanski1983,tianPRB2017}. The temperature dependent Raman spectra shown in Fig.~2a reveal behavior consistent with this, with all observed modes shifting to lower energies as the temperature is increased from 10 to 300~K. A previous report indicated an anomalous decrease in mode energies below $\sim$25~K\cite{wulferding2020}, which was interpreted as evidence for phonon-electron coupling. In an initial run of low temperature measurements we observed similar behavior (see Supplementary Note~5). However, in subsequent measurements we found this anomolous softening was removed by allowing more time for the WP$_{2}$ crystals to come to thermal equilibrium with the cryostat at each temperature.

To extract the quantitative behavior of the phonon energies and linewidths we fit each mode with a Voigt profile. While an asymmetric lineshape can be an indication of phonon-electron coupling\cite{xuNComm17,fanoPR61}, we did not observe this in any WP$_{2}$ mode. Such behavior requires an overlap in energy between an electronic continuum and the discrete phonon state. However the electronic Raman scattering (a quasi-elastic peak)\cite{wulferding2020} appears below the energy of the optical phonons. In Fig.~2b we plot the percent change of each mode's energy and observe that, while most of the modes display changes below 1\%, the lowest energy $A_{1}$ mode shifts by almost 2\%. While this could be a sign of phonon-electron coupling, the temperature dependence of phonon energies contains contributions from multiple sources, so unambiguous interpretation is challenging\cite{tianPRB2017}. We therefore turn to inspection of the phonon linewidths which only contain information related to the decay paths available to the phonon mode\cite{balkanski1983,osterhoudtPRB2019}.

\begin{figure}
\includegraphics{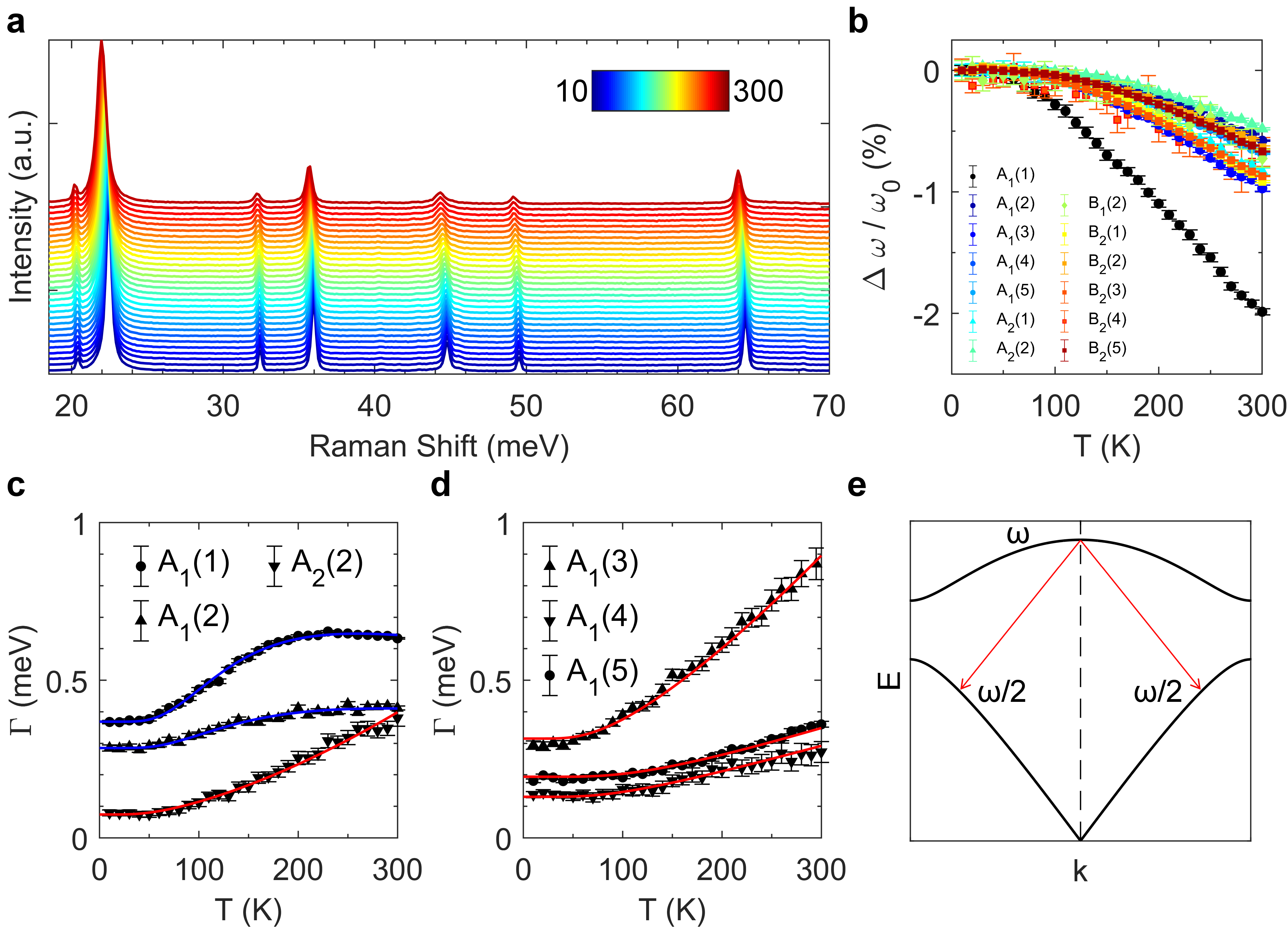}
\caption{(a) Temperature dependent Raman spectra from 10 to 300 K. Spectra are vertically offset. (b) Percent difference of the phonon energies as a function of temperature. The lowest energy $A_{1}$ mode changes by twice the percent of any other mode. (c) Linewidth temperature dependence of the $A_{1}(1)$ and $A_{1}(2)$ modes which have energies 22.4 and 35.8~meV respectively. The $A_{2}(2)$ mode at 32.4~meV has an energy between that of the two $A_{1}$ modes. While both $A_{1}$ modes display an anomalous temperature dependence, the $A_{2}$ mode temperature dependence is that expected from anharmonic phonon decay. Fits in blue and red are those assuming decay into electron-hole pairs and acoustic phonons, respectively. (d) The three higher energy $A_{1}$ modes also display the temperature dependence expected from anharmonic decay. (e) Schematic diagram showing the lowest order anharmonic decay of an optical phonon with energy $\omega$ decay into two acoustic phonons with energy $\omega/2$.}
\label{Fig2}
\end{figure}

The temperature dependent linewidths of the two lowest energy $A_{1}$ modes are plotted in Fig.~2c. At temperatures below $\approx50$~K, the $A_{1}(1)$ mode's linewidth maintains a nearly constant value of 0.36 meV. Above this temperature the linewidth grows rapidly, reaching a maximum value of $\approx0.65$ meV at $\approx200$~K, before slowly decreasing as it approaches room temperature. Though not as dramatically, similar behavior is also seen in the $A_{1}(2)$ linewidth. This type of behavior is distinct from that predicted by phonon decay into acoustic phonons\cite{klemens1966}, as depicted in Fig.~2e. In the Klemens' model, an optical phonon of frequency $\omega$ is assumed to decay into two acoustic phonons of energy $\omega/2$ and momentum $\pm \bvec{k}$ to satisfy both energy and momentum conservation. The inclusion of higher order processes in which an optical phonon decays into three acoustic phonons ($\omega/3$) constitutes the extended Klemens' model\cite{balkanski1983,tianPRB2017}. Due to the bosonic nature of phonons, the temperature dependence in both these models is governed by the Bose-Einstein distribution function $n_{B}(\omega,T)$. Therefore, the linewidth predicted by this model monotonically increases with temperature, a trend that is not displayed by the $A_{1}(1)$ or $A_{1}(2)$ modes. We instead identify their temperature dependence as resulting from decay into electron-hole pairs near the Fermi surface\cite{osterhoudtPRB2019}.

Linewidths indicative of optical phonon decay into electron-hole pairs have previously been observed in the semimetals TaAs\cite{xuNComm17,osterhoudtPRB2019}, NbAs\cite{osterhoudtPRB2019}, MoTe$_{2}$\cite{zhangPRB2019}, graphite\cite{giuraPRB2012,liuC2019} and Cd$_{3}$As$_{2}$\cite{sharafeevPRB2017}. The model developed to describe such behavior assumes an optical phonon with $q = 0$ decays into an electron-hole pair via an interband transition. The decay rate therefore depends on the difference in occupation functions between the electron and hole states\cite{lazzeriPRB06}. Since it is assumed the initial state is below $E_{F}$ and therefore occupied, the model developed for graphene predicts a linewidth monotonically decreasing as temperature is raised due to the occupancy change of the filled and empty states, accounted for by Fermi functions ($n_{F}(\omega,T)$). We modified this account for the nearly parallel spin-orbit split bands of WP$_{2}$: $\Gamma(T) \propto n_{F}(\omega_{a},T) - n_{F}(\omega_{a} + \omega_{ph},T)$, where $\omega_{ph}$ is the energy of the optical phonon, and $\omega_{a}$ captures the distance between $E_{F}$ and the initial state of the electron (see Fig.~3a, additional details in Supplementary Note~7). The inclusion of the $\omega_{a}$ term can result in a non-monotonic temperature dependent behavior because at low temperature the empty initial state prevents the transition. As the temperature is raised the initial state becomes thermally populated and the decay processes turns on. The solid blue lines in Fig.~2c are the results of fitting this model to our measured linewidths, and over the entire measured temperature range it reproduces our experimental data very well. From the fits we find $\omega_{a}$ values of 24.48~meV and 22.64~meV for the $A_{1}(1)$ and $A_{1}(2)$ modes, respectively, consistent with transitions close to the Fermi level.

Having identified the mechanism responsible for the anomalous behavior in the lowest two energy $A_{1}$ modes, we turn to the $A_{2}(2)$ mode whose energy is between the $A_{1}(1)$ and $A_{1}(2)$ modes. As can be seen in Fig.~2c, the linewidth of this mode does not display any anomalous behavior, and instead follows the dependence expected from the model of anharmonic decay. A fit to the data using the extended Klemens' model, plotted as a solid red line, captures the observed trend and confirms this assessment. This observation suggests that the symmetry of the phonon mode plays is central to whether it decays into electron-hole pairs or other phonons. This is further supported by the lack of anomalous behavior in the $B_{1}$ or $B_{2}$ modes (see Supplementary Note~6). However, symmetry is unlikely to be the only determining factor, as evidenced by the linewidths for the three higher energy $A_{1}$ modes plotted in Fig.~2d. Despite sharing the same symmetry, these three modes display behavior consistent with the anharmonic model of decay. Taken together, these observations indicate both symmetry and energy of the optical phonon are factors in determining its primary decay channels. As discussed later, the availability of phonon decay paths, the phonon-electron coupling strength, and momentum conservation  also contribute.

To understand the role of phonon energy, we turn to the available phase space for decay into electron-hole pairs. In Fig.~3a we show the calculated electronic band structure for WP$_{2}$ within $\pm1$~eV of the Fermi level, which agrees with previous work\cite{autesPRL2016,coulterPRB2018}. Along the $\Sigma$ ($\Gamma \to \Sigma_{0}$), $\Delta$ ($\Gamma \to $ Y), and C (Y $ \to $ C$_{0}$) cuts the gaps between neighboring bands appear to be on an energy scale similar to the optical phonons. To confirm this, Fig.~3c shows a low-temperature weighted joint-density of states (JDOS) (see Methods) for vertical ($q=0$) transitions between states within an optical phonon energy ($\pm$70~meV) of $E_{F}$. Available transitions exist across the entire range, with the largest JDOS occurring between 40 and 60~meV. The $A_{1}$ modes that display phonon-electron dominated linewidths have energies of 22.4 and 35.8~meV which have a comparatively smaller electronic JDOS. By contrast, the $A_{1}$ modes that fall in the 40--60~meV range display linewidths governed by anharmonic decay, indicating the availability of electronic states is not the largest contributing factor to the phonon linewidths. As discussed later, we find that momentum conservation is also required to understand the difference between these $A_{1}$ modes.

\begin{figure}
    \centering
    \includegraphics{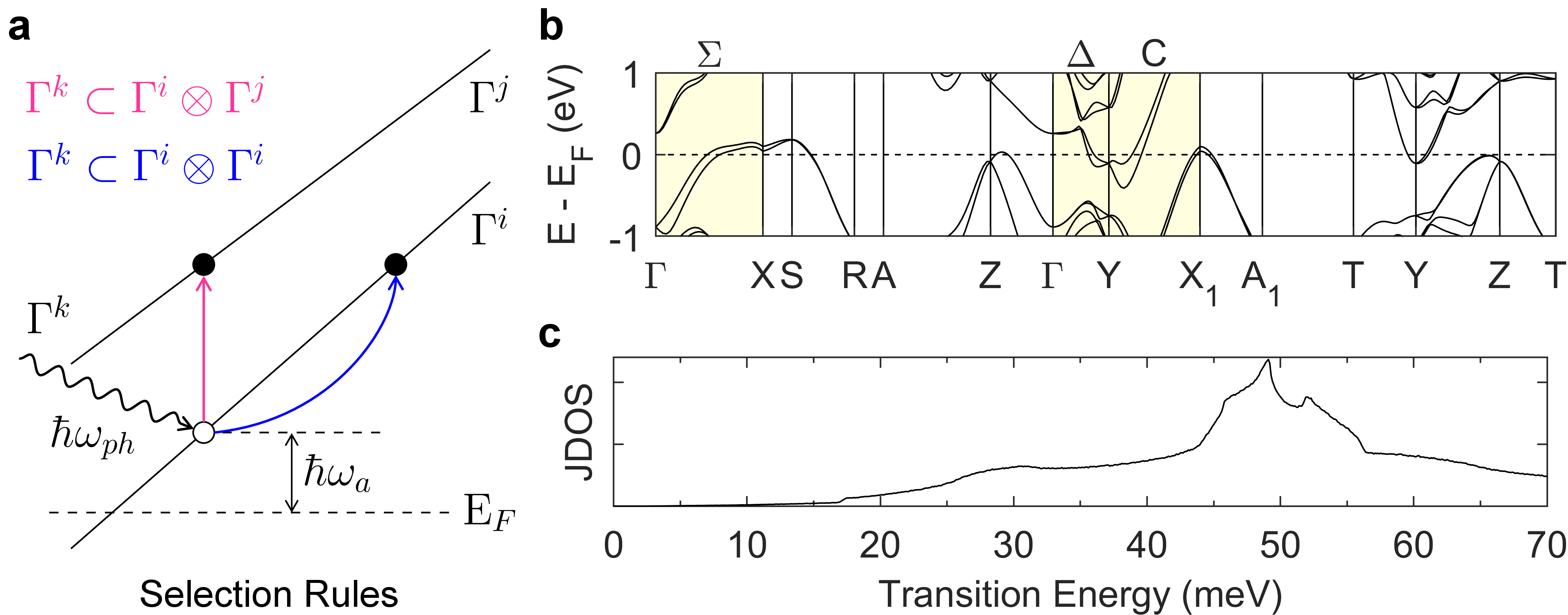}
    \caption{(a) Schematic depicting inter- (pink arrow) and intraband (blue arrow) transitions by optical phonons. An energy $\hbar \omega_{a}$ separates the initial electron state from the Fermi energy E$_{F}$. The electronic bands are labeled by irreducible representations $\Gamma^{i}$ and $\Gamma^{j}$, while the optical phonon of energy $\hbar \omega_{ph}$ has representation $\Gamma^{k}$. $A_{1}$ phonons are  allowed to produce intraband transitions, while they are forbidden from producing interband transitions. (b) Calculated electronic band structure of WP$_{2}$. The shaded regions indicate symmetry lines where scattering by $\Gamma$ point optical phonons is allowed. (c) Weighted joint density of states for vertical ($q = 0)$ transitions between states within $\pm70$~meV of E$_{F}$.}
    \label{fig:3}
\end{figure}

Before addressing momentum conservation, it is instructive to consider the phonon symmetry via the selection rules for transitions between electronic bands. Due to lattice symmetries there are only certain high symmetry points and lines that will allow transitions by $q\approx0$ phonons to occur. The $\Sigma, \Delta,$ and C cuts which both allow these transitions and intersect the Fermi surface are shaded in Fig.~3a. The irreducible representations of the electronic bands along these cuts may be found from compatibility relations\cite{birmanPR1962,chenPR1968} as summarized in Table~\ref{table:comp_rel} and described in detail in the Supplementary information. Due to the inclusion of spin-orbit coupling, all of the electronic bands belong to representations from the double space group, indicated by a bar. By decomposing the direct products of the band representations, we identified the selection rules for phonon decay into both interband and intraband electron-hole pairs. This is schematically depicted in Fig.~3a. Summarizing our results, we find that phonons of $\Gamma^{(2)}, \Gamma^{(3)}$, or $\Gamma^{(4)}$ ($A_{2}, B_{2},$ or $B_{1}$ respectively) symmetry can produce interband transitions, while intraband transitions can result from $\Gamma^{(1)} (A_{1})$, $\Gamma^{(3)}$ and $\Gamma^{(4)}$ phonons. These selection rules show that the $A_{1}(1)$ and $A_{1}(2)$ linewidth behavior observed in Fig.~2c must arise from the creation of electron-hole pairs within the same band. By necessity, this requires phonons of non-zero $\bvec{q}$, a key detail which indicates we must now consider whether momentum conservation can be satisfied under our experimental conditions.

\begin{table}[h]
\centering
\caption{\label{table:comp_rel} Compatibility relations along the highlighted cuts in Fig.~3a.}
\begin{ruledtabular}
\begin{tabular}{ccccc}
    $\Sigma$ & $\Gamma$ & $\Delta$ & Y  &   C  \\
        \hline
        $\Sigma^{(1)}$	&   $\Gamma^{(1)}(A_{1})$	&	$\Delta^{(1)}$	&	$^{(1)}$   &   C$^{(2)}$\\
        $\Sigma^{(2)}$	&	$\Gamma^{(2)}(A_{2})$	&	$\Delta^{(2)}$	&	Y$^{(2)}$   &   C$^{(1)}$\\
        $\Sigma^{(2)}$	&	$\Gamma^{(3)}(B_{2})$	&	$\Delta^{(1)}$	&	Y$^{(3)}$   &   C$^{(1)}$\\
        $\Sigma^{(1)}$	&	$\Gamma^{(4)}(B_{1})$	&	$\Delta^{(2)}$	&	Y$^{(4)}$   &   C$^{(2)}$\\
        $\overline{\Sigma}^{(3)} \oplus \overline{\Sigma}^{(4)}$	&	$\overline{\Gamma}^{(5)}$	&	$\overline{\Delta}^{(3)} \oplus \overline{\Delta}^{(4)}$	&	$\overline{\mathrm{Y}}^{(5)}$   &   $\overline{\mathrm{C}}^{(3)} \oplus \overline{\mathrm{C}}^{(4)}$\\
        \hline\hline
    \end{tabular}
    \end{ruledtabular}
\end{table}

We now consider whether the photons used in our Raman experiments can provide the finite $\bvec{q}$ required for the $A_{1}$ phonons to decay into electron-hole pairs. Near the Fermi energy we assume the electronic bands are approximately linear\cite{yao2019PRL}, with a slope given by the experimentally measured Fermi velocities\cite{goothNComm2018}. The momentum obtained thus depends on the specific electron or hole pocket under consideration (see Supplementary Note~10 for details of the calculation), but for the $A_{1}(1)$ and $A_{1}(2)$ phonons we find $q\approx8.57\times10^{7}\ \text{m}^{-1}$ and $q \approx 1.37\times10^{8}\ \text{m}^{-1}$, respectively. In the backscattering configuration the transferred momentum is found from the relation $q \le 4 \pi n_{i}(\lambda)/\lambda$, where $n_{i}(\lambda)$ is the wavelength dependent index of refraction for the $i$th crystal direction, and $\lambda$ is the wavelength of the optical excitation. We use the calculated index of refraction (see Supplementary Note~11), and find $q \le 9.68\times10^{7}\ \text{m}^{-1}$. We therefore see that the momentum provided by the laser is enough to produce $A_{1}(1)$ phonons capable of intraband transitions. While it does fall short for the $A_{1}(2)$ phonon, we note that uncertainties in the experimental and calculated values, and the finite angles of incidence from our high NA objective could accommodate this discrepancy. Regardless, it is clear that the momentum necessary for an intraband transition by the higher energy $A_{1}$ modes, which we estimate would have a minimum value of $q \approx 1.71\times10^{8}\ \text{m}^{-1}$, is unable to be provided by our laser.

\begin{figure}
    \centering
    \includegraphics[width=\textwidth]{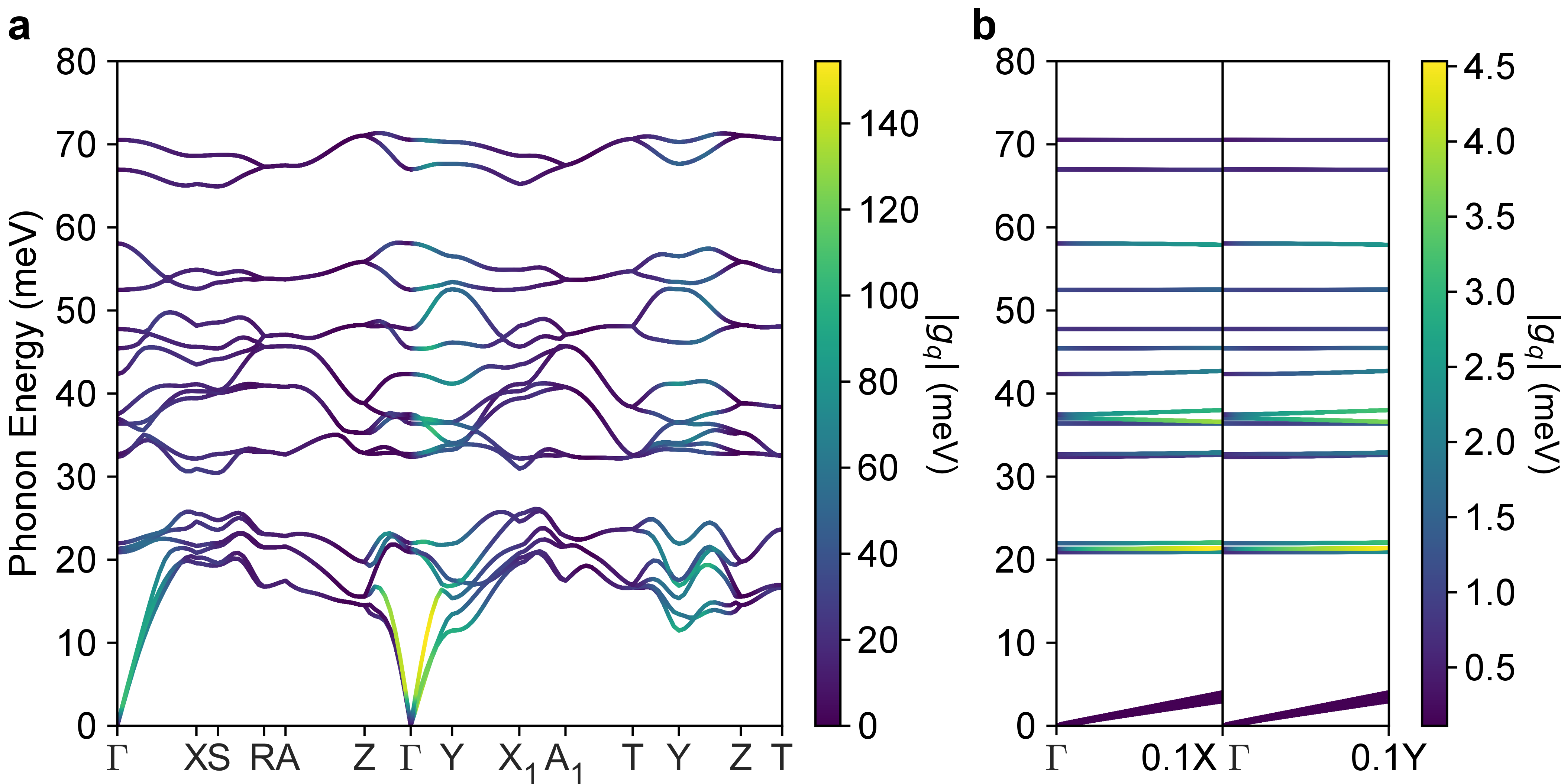}
    \caption{(a) The calculated phonon dispersion of WP$_{2}$. The correction to the imaginary part of the phonon self-energy, is illustrated by the normalized electron-phonon matrix element projected onto each phonon mode ($q,\omega$). The acoustic phonon modes have much stronger coupling than the optical modes. However, when comparing different optical modes, the lower energy modes A$_1$(1) and A$_1$(2) show a significant enhancement of electron-phonon coupling compared to that of the higher energy ones. (b) A close up view of the phonon-electron coupling strength for small $q$ near $\Gamma$, highlighting the largely enhanced coupling strength at finite $q$. Only the optical modes are shown for clarification.}
    \label{fig:4}
\end{figure}

The presence of $A_{2}, B_{1}$, and $B_{2}$ symmetries in the selection rules, combined with the lack of anomalous behavior in the linewidths for phonons of these symmetries, suggests that the phonon-electron coupling strength needs to be considered. Fig.~4a presents the calculated phonon dispersion for WP$_{2}$, with the phonon-electron matrix element shown on each mode. Overall the acoustic branches display greater coupling, while the optical modes have considerable coupling along the $\Delta$ and T to Y to Z cuts, corresponding to phonon wave vectors that connect separate parts of the Fermi surface\cite{osterhoudtPRB2019}. In Fig.~4b we show a close up of the optical phonon dispersion along the $\Sigma$ and $\Delta$ cuts. For $q = 0$ the phonon-electron coupling strength for all modes is very weak, with most modes displaying a $|g|^{2}$ on the order of 10~$\mu$eV. This weak coupling is consistent with the lack of phonon-electron dominated linewiths for modes where interband transitions are symmetry allowed ($A_{2}, B_{1}$, or $ B_{2}$). For finite $q$ we see that for the $A_{1}(1)$ and $A_{1}(2)$ modes, the coupling strength increases by orders of magnitude, while for the higher energy modes the increase in coupling strength is not as dramatic. Noting that this finite $q$ is within our experimentally accessible range, we find these trends in the coupling strength to be consistent with our measurements and interpretation of the phonon linewidths.

The calculated phonon dispersions may also explain why phonon-electron decay dominates over phonon-phonon decay. As shown in Fig.~4a, there is no energy separation between the acoustic branches and the lowest energy optical modes. In addition, the acoustic branches have a very narrow bandwidth, similar to the ``acoustic bunching" discussed for high thermal conductivity materials\cite{broidoPRL13}. These features suggests there is a restricted phase space for decay into acoustic phonons, leading to longer phonon-phonon lifetimes for low energy modes, allowing phonon-electron processes to dominate the linewidth behavior. For the higher energy modes the available phonon-phonon phase space is much larger (Supplementary Note~12), explaining the dominance of phonon-phonon processes for the higher energy $A_{1}$ modes. 

The bunching arguments presented above also apply to the phonon-phonon scattering rates for the acoustic modes.\cite{broidoPRL13} Given the much larger phonon-electron coupling strengths of the acoustic modes (Fig.~4a), we expect that this could lead to a predominance of phonon-electron scattering for these modes. We note that similar conclusions have been drawn from computational works as well\cite{coulterPRB2018}. Scattering by acoustic phonons may be intraband (at low energy and $q$) or interband (higher $q$), which leads to intraband scattering becoming more prevalent at lower temperatures. The acoustic modes are of $A_{1}, B_{1}$, and $B_{2}$ symmetries, and therefore phonon-electron processes are symmetry allowed for two of the branches. If one further considers general $k$-points beyond high-symmetry cuts, the symmetry requirements are relaxed and all three branches may contribute.

By examining the temperature dependence of the optical phonon linewidths of WP$_{2}$ we identified electron-hole decay as the primary path for the lowest energy $A_{1}$ optical phonon modes over a broad temperature range. Through the use of group theory we further identified that such transitions must be intraband, clarifying why only certain Raman modes in this material display this behavior. Combined with first principles calculations, we also explained the importance of the phonon-electron coupling strength and the relative availability of phonon-phonon decay paths. This improved understanding of the differences between phonon induced intraband vs interband scattering is valuable in a material where electron-phonon coupling has been suggested to play a vital role in its macroscopic transport properties. We anticipate that these results could be beneficial to understanding similar phonon behavior that has recently been observed in other Weyl semimetals\cite{osterhoudtPRB2019,zhangPRB2019}.

The much larger phonon-electron coupling strengths calculated for the acoustic modes suggest they may display similar behavior and primarily scatter off of charge carriers, potentially leading to enhancements of the conductivity at low temperatures and contributing to the development of hydrodynamic behavior.\cite{peierls1932,Wiser1984,varnavides2020generalized} Our presented results suggest that enhancing phonon-electron coupling towards novel transport regimes requires lower symmetry systems, strongly bunched acoustic modes, and Fermi surfaces providing a large phase space for such processes. This will require future investigations into the phonon dynamics of topological semimetals utilizing e.g. resonant inelastic x-ray spectroscopy, Brillouin scattering, and new probes of spatially-resolved electrical and lattice transport\cite{Varnavides2019Hydro}.

\section*{Methods}

\subsection{Experimental Details}
Raman spectra were collected in the backscattering configuration using a custom built setup\cite{yaoRSI2016}. The 532~nm light from a frequency doubled Nd:YAG laser was focused by a 100X long-working distance objective to a spot size of $\approx$2 $\mu$m in a Montana Instruments cryostation which enabled access to temperatures from 300 to 10 K. An incident power of <250 $\mu$W was used to achieve satisfactory signal-to-noise ratios, with minimal laser induced heating. This was checked using the Stokes to anti-Stokes ratio at all temperatures where anti-Stokes signal was measurable ($\gtrsim$ 100 K). Polarization dependence was performed by the rotation of a double Fresnel rhomb which acts as a broadband half-waveplate (see Supplementary Information). Any presented spectra have been averaged and had background dark counts subtracted and cosmic rays removed\cite{wangnpj2020}.

\subsection{Computational Details}
Calculations were performed using the \textit{ab initio} package QUANTUM ESPRESSO~\cite{QE-2009,QE-2017} to evaluate electronic band structure, phonon dispersion and electron-phonon interactions, combining density functional theory (DFT) and density functional perturbation theory (DFPT). Fully relativistic optimized norm-conserving Vanderbilt (ONCV) pseudopotentials~\cite{hamann2013optimized} for the PBE exchange-correlation functional~\cite{PBE} were used for both W and P, which allowed us to calculate the electronic structure including spin orbit coupling (SOC). Furthermore, we calculated electron-phonon coupling matrix elements of WP$_2$ with the EPW package~\cite{ponce2016epw}, \textit{via} interpolating the maximally localized Wannier functions~\cite{MarzariVanderbilt1997,Souza2001} on initial coarse $8\times8\times8$ $k$ and $4\times4\times4$ $q$ grids, from which we Fourier transformed back to much finer $60\times60\times80$ $k$ and $q$ grids.

For the weighted joint density of states (JDOS) and refractive index evaluations, separate DFT calculations were carried out using the implementation in JDFTx~\cite{JDFTx}. Fully relativistic ultrasoft pseudopotentials~\cite{dal_corso_pseudopotentials_2014,rappe_optimized_1990} for the PBEsol exchange-correlation functional~\cite{Perdew2008} were used, as well as a uniform $6\times 6\times 8$ $k$ grid for the 6-atom standard primitive unit cell, an energy cutoff of 28 Hartrees, Fermi-Dirac smearing with a 0.01 Hartree width, and a $3\times 3\times 2$ phonon supercell. Maximally localized Wannier functions~\cite{MarzariVanderbilt1997,Souza2001} were similarly obtained to interpolate quantities for Monte Carlo Brillouin zone integration on finer $k$ and $q$ meshes.~\cite{Giustino2007} Each transition contributing to the JDOS at that transition energy is weighted by the factor $n_F(E_{\mathbf{k},n},T)-n_F(E_{\mathbf{k},m},T)$, where $n_F(E,T)$ is the Fermi-Dirac distribution function, and the transition occurs between energies $E_{\mathbf{k},n}$ and $E_{\mathbf{k},m}$ at the momentum $\mathbf{k}$ for two bands indexed by $n$ and $m$. The results presented in Fig.~\ref{fig:3}c are for $T=4$~K, with transitions restricted to be between states within $\pm70$~meV of the Fermi level. The refractive index calculations are discussed in Supplementary Note~11.

\section*{Acknowledgements}
Analysis and measurements performed by G.B.O. and V.M.P.  as well as work done by K.S.B. was supported by the U.S. Department of Energy (DOE), Office of Science, Office of Basic Energy Sciences under Award No. DE-SC0018675. This research used resources of the National Energy Research Scientific Computing Center, a DOE Office of Science User Facility supported by the Office of Science of the U.S. Department of Energy under Contract No. DE-AC02-05CH11231, as well as resources at the Research Computing Group at Harvard University.

G.B.O. and V.M.P. performed the primary measurements, analyzed the data and provided the plots. G.B.O. performed the group theory analysis. C.A.C.G., Y.W. and P.N. performed the first principles calculations. J.G. and C.F. prepared the single crystal samples. G.B.O. and K.S.B. wrote the manuscript with contributions from other authors. P.N. and K.S.B. jointly conceived the project and ideas presented.

%

\end{document}